\documentclass{ws-p8-50x6-00}

\begin{document}

\newcommand{\TeV}{\,{\rm TeV}}
\newcommand{\GeV}{\,{\rm GeV}}
\newcommand{\MeV}{\,{\rm MeV}}
\newcommand{\keV}{\,{\rm keV}}
\newcommand{\eV}{\,{\rm eV}}
\def\ap{\approx}
\def\bea{\begin{eqnarray}}
\def\eea{\end{eqnarray}}
\def\ler{\lesssim}
\def\gtr{\gtrsim}
\def\beq{\begin{equation}}
\def\eeq{\end{equation}}
\def\haf{\frac{1}{2}}
\def\nn{\nonumber}
\def\lpp{\lambda''}
\def\ccg{\cal G}
\def\L{\cal L}
\def\O{\cal O}
\def\R{\cal R}
\def\U{\cal U}
\def\V{\cal V}
\def\W{\cal W}
\def\e{\varepsilon}
\def\slash#1{#1\!\!\!\!\!/}

\setcounter{page}{1}

\title{Brane World in Generalized Gravity
\footnote{Talk given at ``International Workshop on Particle
Physics and the Early Universe'', COSMO2000, Cheju Island, Korea,
September 2000}}

\author{Hyung Do Kim}

\address{Korea Advanced Institute of Science and Technology, Taejon, Korea \\
        and \\
Korea Institute for Advanced Study, Seoul, Korea \\ \tt
hdkim@kias.re.kr}

\date{\today}

\maketitle

\abstracts{ We consider Randall-Sundrum(RS) model in generalized
gravities and see that the localization of gravity happens in
generic situations though its effectiveness depends on the details
of the configuration. It is shown that RS picture is robust
against quantum gravity corrections ($\phi \R$) as long as the
correction is reasonably small. We extend our consideration to the
model of scalar(dilaton) coupled gravity which leads us to the
specific comparison between RS model and inflation models. The
exponential and power law hierarchy in RS model are shown to
correspond to the exponential and power law inflation
respectively.}

\section{Introduction}

Why the gravitational interaction is so weak compared to other
gauge interactions is the main question that made people to
consider the extension of the Standard Model(SM). For several
decades weak scale supersymmetry was believed to be the most
popular explanation for the gauge hierarchy. Recently the presence
of D-brane opened new way of thinking that gravitational
excitations propagate through the full spacetime while gauge
interactions and matter fields are confined on the hypersurface,
so called branes.

The `brane world' scenario changes many conventional viewpoints
toward the problems. First, the weakness of gravity at low energy
is understood by the largeness of the volume of the extra
dimensions \cite{ADD}. This model needs a mechanism of radion
stabilization at large values for its completion which is not easy
without introducing large parameters.

Recently, Randall and Sundrum (RS) \cite{RS1} proposed a new idea
which can explain the gauge hierarchy by localizing gravity on a
`Planck brane' and assuming we are living in a tail (`TeV brane')
of those localized gravity. From now on this will be called RS I
(two brane) model to distinguish it with the single brane setup RS
II \cite{RS2} which has been proposed as an alternative to the
compactification. The localization of gravity yields the physical
mass scale on the TeV brane suppressed by an exponentially small
warp factor. This scenario can be realized with one extra
dimension where $AdS_5$ ends at two boundary 3 branes (Planck and
TeV branes). The 4-D Minkowski solution on the TeV brane can be
achieved only through two exact fine tunings among bulk
cosmological constant and the brane tensions. If the exact
relations among the parameters are not satisfied, we obtain
unstable configurations generically in which the extra dimension
collapses or inflates in addition to the inflation along the 3
spaces parallel to the branes \cite{KK} The solution including
black holes has been obtained in \cite{Kraus}, and the completion
of \cite{KK} has been done in \cite{KSW}. The brane inflationary
solution with fixed extra dimension \cite{Nihei,Kaloper} has been
obtained first. See also \cite{LOW}.

In this paper we investigate the properties of RS model that
remain unchanged when we modify the original simple setup.
Generally quantum corrections alter the simple picture and it is
essential to check whether all nice properties are valid even
after full consideration of quantum corrections which arise
naturally.

\section{Framework}

In this section, we review the general solution generating technique
in the context of scalar coupled gravity mainly following \cite{DFGK}
though the numerical coefficients appearing in the equations
are not the same as in \cite{DFGK}
due to the difference in the metric sign convention or normalization
of the Ricci scalar (or 5-D Planck scale).

Our starting point is the action on $M^4\times S^1/Z_2$ 4with the
metric convention $g_{MN} = (-1,1,1,1,1)$ and
$S=S_{bulk}+S_1+S_2$, whose bulk action is given by \bea
S_{bulk}=\int d^5x \sqrt{-g}\{\frac{1}{2}{\R} -\frac{1}{2} g^{AB}
\partial_A \phi
\partial_B \phi - {\V}(\phi)\},
\eea where $\phi$ is a scalar field {\footnote{Single scalar field
is enough for our purpose since we are interested in the bulk
scalar or dilaton. The generalization including many scalar fields
can be done without any difficulty \cite{DFGK}.} and ${\V}(\phi)$
represents generalized bulk potential including the bulk
cosmological constant and other bulk scalar potentials. The brane
action is \bea S_i=\int d^4x \sqrt{-g_{wall}}\{-
V_i(\phi)+{\L}_{i}\}. \eea Here $V_i(\phi)$ denotes the brane
tension of the $i$-th brane ($i=1,2$) and ${\L}_i$ stands for the
other part of the brane Lagrangian including SM matters which is
assumed to give negligible effects on the spacetime geometry.

The Einstein equation and the equation of motion of $\phi(y)$ are
\bea {\R}_{AB} - \frac{1}{2} g_{AB} {\R} = T_{AB}, \hspace{5mm}
\nabla^2 \phi = \frac{\partial {\cal V}}{\partial \phi}, \nn \eea
where the energy momentum tensors are \bea T_{aa} &=& -\frac{1}{2}
\phi^{\prime 2} - {\cal V(\phi)} -\sum_i V_i (\phi) \delta(y-y_i),
\\ T_{55} &=& \frac{1}{2} \phi^{\prime 2} - {\cal V(\phi)}, \eea
where $A,B,\cdots$ denote 4+1 dimensional indices, and
$a=1,\cdots,3$ denotes the spatial index.

The main interest of this paper is to look at the scaling
behaviors of different theories, and we assume that the radion has
been stabilized. From now on, we focus on the solutions which keep
the 4-D Poincare invariance ($\Lambda_{\rm eff} = 0$). This can be
achieved by tuning the parameter which corresponds to the
cosmological constant problem in 4 dimensional theory. To keep 4-D
Poincare invariance, we set $\phi = \phi (y)$. From the above
conditions, the metric is expressed in a simple form $ds^2 = e^{2
A(y)} \eta_{\mu \nu} dx^{\mu} dx^{\nu} + dy^2$, and we can write
the equation of motion in terms of $\phi(y)$ and $A(y)$. The
integration over $y$ goes from 0 to $L$ since we fixed the scale
factor $g_{55} = 1$.

The Einstein equations can be rewritten in terms of $A(y)$ and
$\phi(y)$ as
\bea &3 A^{\prime \prime} &= - \phi^{\prime 2} -
\sum_i V_i (\phi) \delta (y-y_i), \\ &6 A^{\prime 2} &=
\frac{1}{2} \phi^{\prime 2} - \V (\phi), \\ &\phi^{\prime \prime}
+ 4 A^{\prime} \phi^{\prime} & = \frac{\partial \V}{\partial \phi}
+ \sum_i \frac{\partial V_i (\phi)}{\partial \phi} \delta (y-y_i).
\eea

Though there are three equations, only two of them are
independent. The third equation is derived from the first two
equations. The second equation reminds us of the relation between
the superpotential $W$ and the potential $\V$. If we introduce $W$
and set $\phi^{\prime} = \frac{\partial W}{\partial \phi}$ and
$A^{\prime} = \gamma W$, then from the first equation $\gamma =
-\frac{1}{3}$ is determined. Once we obtain $W(\phi)$ for given
${\V}(\phi)$ from the relation \bea \label{rel} {\V} (\phi) =
\frac{1}{2} (\frac{\partial W}{\partial \phi})^2 - \frac{2}{3}
W^2, \eea the Einstein equations become two first order equations
\cite{CG,DFGK} \bea \label{eqn1} \phi^{\prime} = \frac{\partial
W}{\partial \phi},\hspace{5mm} A^{\prime} = -\frac{1}{3} W \eea
provided the boundary jump conditions \bea \label{bd1} A^{\prime}
|^{y_i+\epsilon}_{y_i-\epsilon} = - \frac{1}{3} V_i (\phi(y_i)),
\hspace{5mm} \phi^{\prime} |^{y_i+\epsilon}_{y_i-\epsilon} =
\frac{\partial V_i (\phi(y_i))}{\partial \phi} \eea are satisfied.
Since we are interested in the scaling behavior of the metric in
the bulk along the extra dimension $y$, we choose $V_i$ such that
the previous jump conditions are always satisfied.

This technique can be applied only if we can find $W(\phi)$
satisfying the relation (\ref{rel}) and is independent of
supersymmetry. `Superpotential' like function $W$ allows us to get
an analytic solution even for some nontrivial potentials.

One of the interesting aspects of RS I model (two brane) is the
radion phenomenology. The Goldberger-Wise(GW) stabilization
mechanism \cite{GW} could generate an exponentially small warp
factor $e^{-2kL}\sim M_W/M_{Pl}\sim 10^{-16}$ without introducing
any large (or small) numbers in the model.

Csaki et. al.\cite{CGRT}, as well as Goldberger and Wise\cite{GW2},
noted that the radion kinetic term arising from the Ricci scalar $\R$
is exponentially small ($\sim e^{-4kL}$) as a result of
nontrivial cancellations between various
terms of order one. Compared to the case of order one kinetic
term, this gives  dramatically different radion phenomenology.
Radion mass is essentially of the order of weak scale
and radion couplings to the standard model fields are
of the order of weak interaction strength.
If the kinetic term were of order one, one would have
radion mass $m_b^2 \sim (M^2_W/M^2_{Pl}) M_W^2$ and  radion couplings
of gravitational strength.

However, this peculiar phenomenon is due to the choice of unnatural gauge
and disappears when we choose more natural gauge \cite{CGR}.
The actual wave function of radion is concentrated on TeV brane,
and it is very reasonable to have a weak scale radion mass
since the mass is not suppressed by large volume factor.

In the following sections, we see how the radion potential changes
when the quantum corrections are added.

\section{Conformal transformation}
Now we have a tool which is very useful for the system
with Einstein-Hilbert action and general kinds of scalar potential.
However,
we need more than this to consider $\phi \R$ correction.
It is well known \cite{Maeda} that generalized action of the following form
can be transformed to the Einstein frame
by conformal transformation
\beq
S_{bulk}=\int d^5x
\sqrt{-g}\{\frac{1}{2}f(\phi) {\R}
-\frac{1}{2} g^{AB} \partial_A \phi
\partial_B \phi - {\V}(\phi)\}.
\eeq
The conformal transformation
\bea
\hat{g}_{MN} = e^{2\omega} g_{MN}
\eea
with $\omega = \frac{1}{3} \log (f(\phi))$
brings us to the Einstein-Hilbert action
\beq
S_{bulk}=\int d^5x
\sqrt{-\hat{g}}\{\frac{1}{2}{\R}(\hat{g})
-\frac{1}{2} {\hat{g}}^{AB} \partial_A \phi
\partial_B \Phi - {\U}(\Phi)\}
\eeq where the potential in Einstein frame is \bea
\label{conformalV} {\U}(\Phi) = (f(\phi))^{-\frac{5}{3}} {\V}
(\phi). \eea and the canonically normalized field $\Phi$ which has
a relation with $\phi$ is \cite{Maeda} \bea \label{conformal} \Phi
= \int d\phi (\frac{f(\phi) +
{\frac{8}{3}}(\frac{df(\phi)}{d\phi})^2}
{f^2(\phi)})^{\frac{1}{2}}. \eea All the cases appearing in the
following are analyzed in two steps based on these tools. First,
take a conformal transformation to Einstein frame. Second, find
$W$ and solve the first order differential equations.

\subsection{Small $\phi \R$ correction}

Let's consider $\phi \R$ correction first. When scalar fields are
present, the effective action is expected to contain $\phi \R$
term which is induced by quantum corrections \beq S_{bulk}=\int
d^5x \sqrt{-g}\{\frac{1}{2}(1+\e \phi^2) {\R} -\frac{1}{2} g^{AB}
\partial_A \phi
\partial_B \phi - {\V}(\phi)\}.
\eeq It is reasonable to assume that the coefficient of the
induced term is small $\e \ll 1$ (e. g. $\e \sim 1/100)$ since it
usually contains a loop suppression factor. We can take the
conformal transformations such that \beq S_{bulk}=\int d^5x
\sqrt{-g}\{\frac{1}{2}{\R} -\frac{1}{2} g^{AB}
\partial_A \Phi
\partial_B \Phi - {\U}(\Phi)\},
\eeq where $\Phi$ and ${\U}(\Phi)$ are determined by
(\ref{conformal}), (\ref{conformalV}) as \bea \label{cphi} \Phi  =
\phi - \e \phi^3 + {\O}(\e^2), \hspace{5mm} {\U}(\Phi) & =
&(1-\frac{5}{3} \e \phi^2) {\V}(\phi) + {\O}(\e^2). \eea

The action on the brane has not been specified, and we assume that
the correction on the brane is such that the relation (\ref{bd1})
is satisfied to keep the brane configuration static.

Before considering quantum correction, let's review the back
reaction of GW scalar field. When there was no potential for the
scalar $\phi$ and ${\V} = \Lambda = - 24 k^2$, we recover the RS
model with $W = 6k$ and $A = -2ky$. To check whether $\phi \R$
correction destabilize the GW mechanism, we should consider
massive bulk scalar at first. Massive bulk scalar changes the
potential and now ${\V} = \Lambda + m_{\phi}^2 \phi^2 = -24k^2 (1
+ \eta \phi^2)$ in which we introduce new parameters, $k$
representing the bulk cosmological constant and $\eta$, the ratio
of the scalar mass to the bulk cosmological constant. Now $W = 6k
(1+ \frac{1}{2}\eta \phi^2) + {\O} (\eta^2)$ where $\eta \ll 1$
when the mass is small compared to the bulk cosmological constant
which is the case for stabilizing the radion at order 10 value
($kL \approx 37$). The bulk geometry is not pure $AdS_5$ due to
the back reaction of the bulk scalar \cite{DFGK}, and we obtain
the metric $g_{\mu\nu} = e^{2A} \eta_{\mu\nu}$ from eq.
(\ref{eqn1}) as \bea \label{backreaction} A = - 2k
(1+\frac{1}{2}\eta \phi_0^2 ) y + {\O} (\eta^2), \eea where
$\phi_0$ is the value of $\phi$ at $y=0$. Though there is a
nontrivial(not linear) $y$ dependence due to the back reaction
appearing at ${\O} (\eta^2)$, we neglect it in this section since
first order approximation is enough if $\eta$ is small enough
($\eta \ll 1$).

Let's back to the quantum correction $\phi \R$. The above eq.
(\ref{cphi}) then become after conformal transformation \bea {\U}
(\Phi) & = & -24 k^2 (1 + (\eta-\frac{5}{3} \e) \phi^2)
+{\O}(\e^2, \eta^2, \e \eta), \\ {\W} (\Phi) & = & 6k ( 1+
\frac{1}{2}(\eta-\frac{5}{3} \e) \phi^2) +{\O}(\e^2, \eta^2, \e
\eta). \eea These equations give us very simple interpretations of
the $\phi \R$ correction as the correction of the bulk scalar
masses. The change in the geometry is given as the same way as in
eq. (\ref{backreaction}) with the effective mass by replacing
$\eta \rightarrow (\eta-\frac{5}{3} \e)$. This opens new
possibility of generating bulk scalar masses through quantum
gravity corrections. Scalar mass can be obtained from this quantum
correction even though we started from the setup with a massless
bulk scalar.

All the pictures in RS I (with two branes)
, e.g., the generation of huge hierarchy between Planck brane
and TeV brane, Goldberger-Wise stabilization mechanism,
remain the same
as long as $\epsilon$ is small enough,
for instance, $\epsilon \le 10^{-2}$.
%for instance, $\epsilon \lesssim 10^{-2}$.
The effects of $\phi \R$ is to modify the scalar mass. Though it
is possible to imagine that scalar mass is originated from quantum
gravity correction, we have kept in mind that the small scalar
mass is based on the symmetry which is slightly broken, e.g., bulk
supersymmetry. In that case the quantum correction to the scalar
mass is proportional to the tree level scalar mass itself and is
loop suppressed, $\e \sim \frac{1}{100} \eta \ll \eta$. Therefore
$\phi \R$ correction is not harmful to the radion potential, and
the RS solution to the gauge hierarchy can be kept stable against
small $\phi \R$ correction.

\subsection{Brans-Dicke Theory}

In this section, we consider different types of theories
which have entirely different properties than RS model.
String theory has a dilaton which couples directly with scalar curvature
and give Brans-Dicke (BD) type theory as its low energy effective theory
This kind of theory shows very different behavior than
the RS model. Already it has been shown that exponential and power law
hierarchy are obtained in the usual supergravity and the gauged supergravity
respectively in the framework of strongly coupled heterotic string
\cite{Kehagias}.

Now all the features of RS scenario
change if we consider BD type interactions
between scalar fields and gravity.
To see the qualitative features,
we consider the following action
\bea
S_{bulk}=\int d^5x
\sqrt{-g}\{\frac{1}{2} \e \phi^2 {\R}
-\frac{1}{2} g^{AB} \partial_A \phi
\partial_B \phi - {\V}(\phi)\}.
\eea
Actually typical form of BD theory is
\bea
S_{bulk}=\int d^5x
\sqrt{-g}\{\frac{1}{2} \tilde{\phi} {\R}
- \frac{1}{2} g^{AB} \frac{\omega}{\tilde{\phi}} \partial_A \tilde{\phi}
\partial_B \tilde{\phi} - {\V}(\tilde{\phi})\},
\eea but this is equivalent to the previous one with the relation
$\e = \frac{1}{4\omega}$. The equivalence relation can be easily
checked by changing the BD kinetic term to the canonical form.

Conformal transformation brings the action into
\beq
S_{bulk}=\int d^5x
\sqrt{-g}\{\frac{1}{2}{\R}
-\frac{1}{2} g^{AB} \partial_A \Phi
\partial_B \Phi - {\U}(\Phi)\},
\eeq where $\Phi$ and ${\U}(\Phi)$ are determined from $\phi$ and
${\V}(\phi)$ using eq. (\ref{conformal}) and (\ref{conformalV}).
Now the situation is entirely different from the previous case.
$\Phi$ and $\phi$ are not related linearly, \bea \Phi &=&
\sqrt{\frac{1+\frac{32}{3} \e}{\e}} \log (\phi/\phi_0) = a \log
(\phi/\phi_0), \eea and the potential has exponential dependence
\bea {\U}(\Phi) & = & \e^{-\frac{5}{3}} \phi_0^{-\frac{10}{3}}
e^{-\frac{10}{5a}\Phi} {\V}(\phi(\Phi)). \eea When the bulk
cosmological constant is dominant, \bea W(\phi)  =  6k,
\hspace{5mm} {\V}(\phi)= -24k^2, \eea we get \bea {\W}(\Phi) & = &
6k \e^{-\frac{5}{6}} \phi_0^{-\frac{5}{3}} e^{-\frac{5}{3a}\Phi} =
c e^{-\frac{5}{3a}\Phi} \eea and can solve the differential
equation \bea A & = & \frac{3 a^2}{25} \log ( 1 - \frac{25}{9 a^2}
c y )
 = \frac{3+32\e}{25\e} \log ( 1 - \frac{25\e}{9+96\e} c y )
\eea where $c = {\W} (y=0)$ is the integration constant. Now the
metric is \bea \label{power} g_{\mu\nu} = e^{2A} \eta_{\mu\nu} = (
1 - \frac{25\e}{9+96\e} c y) ^{\frac{6+94\e}{25\e}} \eta_{\mu\nu},
\eea and it shows the power law dependence along $y$ which becomes
singular at finite $y$ ($y_c = \frac{9+96\e}{25\e c}$).
Consideration of exponential potential rather than power law
potential gives the same result, and this corresponds to the
self-tuning brane models having singularity when we cut off the
bulk at the singularity \cite{KSS,ADKS}. Cutting the bulk before
the singularity occurs gives the usual string settings like
Horava-Witten \cite{HW}. The potential generating power law
hierarchy has been studied in \cite{BHLLM} independently.

By taking the limit $\e \rightarrow 0$ with $\e \phi^2$ fixed, we
can recover the RS model as $g_{\mu\nu} = \e^{2A} \eta_{\mu\nu} =
\e^{-4c y} \eta_{\mu\nu}$. In this limit $\phi$ is frozen since
the kinetic term becomes huge, and the system recovers
Einstein-Hilbert action. Even for a tiny but nonzero $\e$ ($\e \ll
1$), the metric (\ref{power}) shows the presence of singularity at
$y_c \simeq \frac{9}{25 c \e}$. At any rate, we can generate the
hierarchy $10^{-32}$ with $\e \simeq 1/120$ and putting TeV brane
at $y = 0.9 y_c \simeq 40 c \sim 40$ for order one $c$. If we put
TeV brane at $y = 0.99 y_c$, we need $\e \simeq 1/50$ and $y
\simeq 20 c \sim 20$.

\section{Randall-Sundrum vs. Inflation}

Understanding of RS geometry can be easily done by thinking $y$ as
time of the inflation models. RS model itself corresponds to the
usual inflation models in which the expansion is exponential, and
scalar(dilaton) coupled theory gives rise to the power law
hierarchy as in the case of extended inflation model in which BD
theory gives power law inflation. More concrete relations can be
found in the recent paper \cite{HDK} with a table summarizing it.
This analogy shows that singularities away from the brane is
inevitable since this corresponds to the initial singularity in
inflation models. RS II (single brane) model has AdS geometry even
far away from the brane with decreasing warp factor. However, an
analogy with inflation models shows that asymptotically AdS
geometry with decreasing warp factor is not an attractor of the
system. This is the opposite case to the inflation in which
asymptotically dS is an attractor. The difference is due to the
fact that the direction we are considering is opposite with each
other. If we consider time reversal direction, asymptotically dS
is not an attractor and generally singularity is developed since
the connection term in the scalar field equation of motion acts
like an anti-friction. This singularity has already been observed
in the study of 5-D supergravity inspired by AdS/CFT
correspondence. The paper \cite{HDK} addresses the question of
which types of singularities are harmless. In other words the
criterion on which types of the singularities are expected to be
resolved is suggested.

\section{Conclusion}

We have studied general features of RS model by considering
several quantum corrections and checked that RS model is robust
against small quantum corrections. Though exponentially suppressed
potential looks suspicious to be stable against possible dangerous
corrections, it turned out that nothing is harmful as far as we
are concerned on a theory whose systematic expansion is possible.
We considered $\phi \R$ term and showed that it acts like changing
the bulk scalar mass. Therefore, once the mass is small, the
correction is further suppressed and the radion potential remains
stable. Also we obtained the power law $y$ dependence of the
metric for the BD type generalization of RS model. Exponential law
$y$ dependence of the warp factor which is important to localize
the gravity can be realized by freezing the dilaton such that
there is no scalar that couples directly to $\R$. Otherwise, the
singularity appears and we can not have RS II (single brane model)
as an alternative of compactification. Finally we stated several
correspondences between RS model and inflation model and checked
that asymptotically $AdS$ do not come as an attractor of the
system which is necessary for the localization of gravity. This is
opposite to the slow roll condition of the inflation models where
the condition is guaranteed by appearing as an attractor. The
singularity entering in general setup with scalar fields is
inevitable in RS II (single brane setup). Nonetheless, RS I model
(two brane) remains as very robust one at least within our
consideration since the consideration of quantum effects does not
alter its characteristic features.

\section*{Acknowledgements}

HK thanks to theory group of Univ. of California at Santa Cruz
including T. Banks, M. Dine, M. Graesser, H. Haber and L. Motl for
the hospitality during the visit. HK also indebted to K. Choi for
many valuable comments and discussions.

\end{document}